\documentclass[twocolumn,showpacs,preprintnumbers]{revtex4}
\usepackage{amsmath}
\usepackage{amssymb}
\usepackage{graphicx}
\usepackage{dcolumn}
\usepackage{bm}

\begin{document}

\title{Modification of spin mixing of spinor BEC by cavity QED coupling}
\author{Ma Luo, Chengguang Bao, Zhibing Li\footnote{%
Corresponding author: stslzb@mail.sysu.edu.cn}}
\affiliation{State Key Laboratory of Optoelectronic Materials and Technologies \\
School of Physics and Engineering \\
Sun Yat-Sen University, Guangzhou, 510275, P.R. China}

\begin{abstract}
Dressed states of spinor Bose-Einstein condensates of spin-1 atoms
coupling with optical cavity modes with far off resonance frequency
are investigated. The exact solution of time evolution of population
of spin component is derived, and the numerical result shows that
the evolution is different from spin mixing without the coupling.
Due to the coupling with the atoms, the photon state also evolute to
different optical cavity modes.
\end{abstract}


\maketitle

\section{Introduction}
Cavity Quantum Electrodynamics (cavity QED)
\cite{01boca04,02manuz05} describes the coupling between atom and
confined cavity mode of electromagnetic field by coherent
interaction. It gives a great chance to reveal the nature of
coherent interaction between matter and field, and to various
application such as quantum information and quantum state
engineering. The atom can be described by two-level system, and the
cavity mode is described as quantized light field. The coupling
between two-level system and quantized light field is described by
the Jaynes-Cummings model \cite{03jaynes63}, which describe the
coupling strength as the speed of energy transfer between two
subsystem (here the excitation energy of the atom is close to energy
level of cavity mode, which is called near resonance). When the
energy level distribution of the atom is more complicated,
three-level system or other multi-level system is needed to describe
the atom. But the description of coupling is similar. Some
multi-level system give interesting phenomenon such as
electromagnetical induced transparent \cite{04peter06}.

When there are more than one atom, the system can be described by
Tavis-Cummings model \cite{05tavis68}, where N atoms are assumed to
couple to the same cavity mode with identical coupling strength.
This lead to enhance of the coupling strength to $\sqrt{N}$ time.
When the temperature of the system is lower than critical
temperature, the N atoms couple with each other coherently and form
Bose-Einstein condensate (BEC). In this case, all atoms have the
same quantum state, which ensure the same coupling strength with
cavity mode. As a result, the Tavis-Cummings model is very
appropriate for this system. A simpler model is to describe the BEC
of the N atoms by mean field theory. A dressed BEC state
\cite{06elena98} and a dark state \cite{07lee99} is investigated by
the mean field theory description. When the energy level of cavity
mode is much different from excitation energy of the atoms (far off
resonance), the coupling can be described by effective Hamiltonian
\cite{06elena98}, which is given by solving the Schrodinger equation
of single atom and cavity mode. The dynamic property of dress BEC
state of far off resonance is also investigated under mean field
theory method \cite{06elena98}.

When the N atoms is trapped in an total optical trap, the spin
freedom of the atoms is librated. When temperature is low enough,
spinor BEC is form \cite{08ho98,stenger98}, which has much potential
in application of quantum computer and quantum information
\cite{sorensen01}. Spinor BEC can be described by either mean field
theory or analytical express of the quantum state. A former work
about fractional parentage coefficients (FPCs) gives a chance to
describe spinor BEC of spin-1 atoms in a simpler way
\cite{09bao04,10bao05}. One of the most interesting phenomenon of
spinor BEC is spin mixing, where the average atomic number of each
spin component evolute versus time \cite{11chang05,12law98}. The
pattern of time evolution is found to be decided by symmetric
property and initial condition of the system \cite{ma08}. If the
spinor BEC is placed inside a far off resonant cavity, the atoms
will couple to the cavity mode, which change the symmetric of the
system as well as the effective coupling strength between atoms. As
a result, the evolution of spin component is supposed to be
modified. In this paper, we will deduce the effective Hamiltonian to
describe the far off resonant coupling of spinor BEC and cavity
mode, and figure the dressed state of spinor BEC which is eigenstate
of the Hamiltonian. And then we investigate the modification of spin
mixing by the coupling. The spinor BEC system is described by FPCs,
which can give an analytical solution to the problem \cite{10bao05}.

\section{Description of Dressed Spinor BEC State}

The system we are going to investigate consist of $N$ $^{87}Rb$
atoms which are optically trapped to form a spinor BEC, and located
inside a high-Q optical cavity. Assume that there are only two
optical modes of frequency $\omega_{c}$, which are left and right
circular polarization optical mode in the cavity. The two optical
modes have opposite angular momentum $\hbar$ along $\hat{z}$
direction, which are indexed by $\mu=\pm 1$. The ground state of
$^{87}Rb$ is $5^{2}S_{1/2},F=1$, and the lowest energy state of
excited state of $D1$ line is $5^{2}P_{1/2},F=1$ \cite{hans97}. When
$\omega_{c}$ is smaller than and at the same scale of the frequency
of $5^{2}S_{1/2},F=1\longrightarrow5^{2}P_{1/2},F=1$ transition
$\omega_{e}$, we can neglect other transition. As a result, the
non-coupling Hamiltonian of the N atoms with spin dependent
interaction with the optical frequency removed is
\begin{eqnarray}
H_{0}&=&\sum_{i=1}^{N}(\frac{\mathbf{p}_{i}^{2}}{2m}+U(\mathbf{r}_{i})+\sum_{\sigma=1,0,-1}\hbar\delta|e_{i},\sigma\rangle\langle
e_{i},\sigma|)\notag \\
&+&\sum_{i\neq j}V_{i,j}
\end{eqnarray}
where $\mathbf{p}_{i}$ is momentum of the ith atom, $m$ is atomic
mass, $U(\mathbf{r}_{i})$ is potential of external optical trap;
$\delta=\omega_{e}-\omega_{c}$, $|e_{i},\sigma\rangle$ is the
$5^{2}P_{1/2},F=1$ state of the ith atom with spin component
$\sigma$ ($|g_{i},\sigma\rangle$ is the $5^{2}S_{1/2},F=1$ state of
the ith atom with spin component $\sigma$);
$V_{i,j}=\delta(\mathbf{r}_{i}-\mathbf{r}_{j})(c_{0}+c_{2}\mathbf{F}_{i}\cdot\mathbf{F}_{j})$
is spin dependent interaction between atoms, with $\mathbf{F}$ being
spin operator and $c_{0,2}$ being interaction constant
\cite{08ho98}.

Since the Q factor of the optical cavity is very high, and
temperature is very close to zero, the decay of photon and
non-coherent interaction between atoms and photon can be neglected.
According to the selection rule, if the atom absorb (emit) one
photon from(to) the cavity, the allowed transitions have
$\Delta\sigma=\pm 1$. Therefore, the Hamiltonian of coupling between
atom and cavity mode is given by
\begin{eqnarray}
H_{C}=\sum_{i=1}^{N}\hbar\Omega_{0}(|e_{i},0\rangle\langle
g_{i},-1|\hat{a}_{1}+|e_{i},1\rangle\langle
g_{i},0|\hat{a}_{1}\notag \\
+|e_{i},-1\rangle\langle g_{i},0|\hat{a}_{-1}+|e_{i},0\rangle\langle
g_{i},1|\hat{a}_{-1}+c.c)
\end{eqnarray}
where $\hat{a}_{\mu}$ and $\hat{a}_{\mu}^{+}$ are annihilation and
creation operators of optical modes of polarization $\mu=\pm 1$,
$\Omega_{0}=dE/\hbar$ is the strength of the atom-field coupling.
$d$ is the atomic dipole-matrix element, and
$E=\sqrt{\hbar\omega_{c}/2\epsilon_{0}V}$ is electric field of
single photon state, $V$ being mode volume. In far off resonance
region, we have $\delta>>\Omega_{0}$.

Using similar process in reference \cite{06elena98}, we can get the
effective Hamiltonian of the coupling system without involving the
atomic excited state
\begin{eqnarray}
H_{eff}=&&\sum_{i=1}^{N}[\frac{\mathbf{p}_{i}^{2}}{2m}+U(\mathbf{r}_{i})\notag
\\
+&&\frac{\hbar\Omega_{0}^{2}}{\delta}(-|g_{i},-1\rangle\langle
g_{i},-1|a_{1}^{+}a_{1}-|g_{i},1\rangle\langle
g_{i},1|a_{-1}^{+}a_{-1}\notag
\\
&&+|g_{i},-1\rangle\langle
g_{i},1|a_{1}^{+}a_{-1}+|g_{i},1\rangle\langle
g_{i},-1|a_{-1}^{+}a_{1})]\notag
\\+&&\sum_{i\neq j}V_{i,j}\label{e3}
\end{eqnarray}
It is obvious that the last two single particle terms are given by
the process that the atom absorb one photon from one mode and emit
one photon to the another mode. Note that the Hamiltonian (\ref{e3})
conserve the total angular momentum at $z$ direction $\mathcal
{M}\hbar$ and total number of photons $n=n_{1}+n_{-1}$. In order to
obtain the eigen state of the Hamiltonian, we first give the eigen
state of the Hamiltonian without coupling between atom and optical
mode ($\Omega_{0}=0$), which is direct product of eigen state of the
N atoms and eigen state of optical mode. And then calculate the
matrix element of the Hamiltonian with coupling ($\Omega_{0}\neq0$)
under the basic states given by the direct product states. The
eigenstate of the Hamiltonian is obtained by diagonalizing the
matrix.

The direct product states is given as
\begin{equation}
|S,M;n_{1},n_{-1}\rangle=|S,M\rangle_{N}\otimes|n_{1},n_{-1}\rangle
\label{e4}
\end{equation}
where $|S,M\rangle_{N}$ is eigenstate of the N atoms with total spin
$S$ and $z$ component of total spin $M$, $|n_{1},n_{-1}\rangle$ is
optical state with $n_{1}$ ($n_{-1}$) photons in $\mu=1(-1)$
polarization state. The $z$ direction angular momentum of the whole
system is $\mathcal {M}\hbar=(M+n_{1}-n_{-1})\hbar$, and eigen
energy of the states is $E_{S,n_{1},n_{-1}}=S(S+1)J$, where $J$ is a
constant decided by spin dependent interaction $c_{2}$. The atomic
eigenstate is given as
\begin{equation}
|S,M\rangle_{N}=\vartheta^{N}_{S,M}\prod_{i=1}^{N}\phi(\mathbf{r}_{i})
\end{equation}
where $\vartheta^{N}_{S,M}$ is the total symmetric spin state of N
atoms, $\phi(\mathbf{r})$ is the single particle spatial wave
function. Note that we use second quantized Hamiltonian to describe
optical mode, while use first quantized Hamiltonian to describe the
N atoms. The total symmetric property of the N Bosonic atoms is
included in the total symmetric atomic wave function. The total
symmetric spin state can be expanded as
\begin{equation}
\vartheta^{N}_{S,M}=a^{N}_{S}[\chi(i)\vartheta^{N-1}_{S+1}]_{S,M}+b^{N}_{S}[\chi(i)\vartheta^{N-1}_{S-1}]_{S,M}
\label{e6}
\end{equation}
where $a_{S}^{N}=\{[1+(-1)^{N-S}](N-S)(S+1)/[2N(2S+1)]\}^{1/2}$ and
$b_{S}^{N}=\{[1+(-1)^{N-S}]S(N+S+1)/[2N(2S+1)]\}^{1/2}$ are FPCs
given in reference \cite{10bao05}, $\chi(i)$ is spin state of the
ith spin-1 ground state atom, which is couple to total symmetric
spin state of the other $N-1$ atoms by Clebsch-Gordan coefficient.

The eigenstates with quantum number ($\mathcal {M}\hbar,n$) can be
expanded by the direct product states (\ref{e4}) with the same value
of $\mathcal {M}\hbar$ and $n=n_{1}+n_{-1}$. Since we are interested
in the quantum effect of the system, we focus on one photon case,
i.e. $n_{1}+n_{-1}=1$, but the extension to more photon case is
straight forward. When there is only one photon, only the direct
product states with quantum number
$(M=\mathcal{M}-1,n_{1}=1,n_{-1}=0)$ or
$(M=\mathcal{M}+1,n_{1}=0,n_{-1}=1)$ can be basic states. As a
result, the eigenstate can be expressed as
\begin{eqnarray}
|\mathcal{M},\nu\rangle=\sum_{S>=\mathcal{M}-1}
c^{\nu}_{S,\mathcal{M}-1}|S,\mathcal{M}-1;1,0\rangle \notag
\\
+\sum_{S>=\mathcal{M}+1}
c^{\nu}_{S,\mathcal{M}+1}|S,\mathcal{M}+1;0,1\rangle
\end{eqnarray}
where $\nu$ is the index of eigenstate,
$c^{\nu}_{S,\mathcal{M}\pm1}$ is expansion coefficients, and the
summation satisfy $N-S$ is even. Making use of equation (\ref{e6}),
the matrix elements are obtained as
\begin{eqnarray}
\langle
S,\mathcal{M}-\mu;\{\mu\}|H_{eff}|S,\mathcal{M}-\mu;\{\mu\}\rangle
=S(S+1)J\notag
\\-\frac{\hbar\Omega_{0}^{2}N}{\delta}[(A^{N}_{S,\mathcal{M}-\mu,-\mu})^{2}+(B^{N}_{S,\mathcal{M}-\mu,-\mu})^{2}]
\end{eqnarray}
\begin{eqnarray}
\langle
S+2,\mathcal{M}-\mu;\{\mu\}|H_{eff}|S,\mathcal{M}-\mu;\{\mu\}\rangle
=\notag
\\-\frac{\hbar\Omega_{0}^{2}N}{\delta}A^{N}_{S,\mathcal{M}-\mu,-\mu}B^{N}_{S+2,\mathcal{M}-\mu,-\mu}
\end{eqnarray}
\begin{eqnarray}
\langle
S-2,\mathcal{M}-\mu;\{\mu\}|H_{eff}|S,\mathcal{M}-\mu;\{\mu\}\rangle
=\notag
\\-\frac{\hbar\Omega_{0}^{2}N}{\delta}A^{N}_{S-2,\mathcal{M}-\mu,-\mu}B^{N}_{S,\mathcal{M}-\mu,-\mu}
\end{eqnarray}
\begin{eqnarray}
\langle
S,\mathcal{M}+\mu;\{-\mu\}|H_{eff}|S,\mathcal{M}-\mu;\{\mu\}\rangle
=\frac{\hbar\Omega_{0}^{2}N}{\delta}\notag
\\
(A^{N}_{S,\mathcal{M}+\mu,\mu}
A^{N}_{S,\mathcal{M}-\mu,-\mu}+B^{N}_{S,\mathcal{M}+\mu,\mu}
B^{N}_{S,\mathcal{M}-\mu,-\mu})
\end{eqnarray}
\begin{eqnarray}
\langle
S+2,\mathcal{M}+\mu;\{-\mu\}|H_{eff}|S,\mathcal{M}-\mu;\{\mu\}\rangle
=\notag
\\\frac{\hbar\Omega_{0}^{2}N}{\delta}A^{N}_{S,\mathcal{M}-\mu,-\mu}B^{N}_{S+2,\mathcal{M}+\mu,\mu}
\end{eqnarray}
\begin{eqnarray}
\langle
S-2,\mathcal{M}+\mu;\{-\mu\}|H_{eff}|S,\mathcal{M}-\mu;\{\mu\}\rangle
=\notag
\\\frac{\hbar\Omega_{0}^{2}N}{\delta}A^{N}_{S-2,\mathcal{M}+\mu,\mu}B^{N}_{S,\mathcal{M}-\mu,-\mu}
\end{eqnarray}
where $\{\mu=1(-1)\}$ stand for optical state $|1,0\rangle$
($|0,1\rangle$), $A^{N}_{S,M,\mu}=a^{N}_{S}C^{SM}_{S+1,1,M-\mu,\mu}$
and $B^{N}_{S,M,\mu}=b^{N}_{S}C^{SM}_{S-1,1,M-\mu,\mu}$ with
$C^{S,M}_{S_{1},S_{2},M_{1},M_{2}}$ being Clebsch-Gordan coefficient
\cite{varshalovich88}. The eigenstate $|\mathcal{M},\nu\rangle$ is
the dressed spinor BEC state.

\section{Modified Spin Mixing}

In experiment of spin mixing, the initial atomic state is a Fock
state $|N_{0},M\rangle$ that each spin component has fix number of
atoms. $N_{0}$ is atomic number of spin component 0, $M\hbar$ is $z$
direction angular momentum. Atomic number of spin component 1 and -1
can be expressed by $N_{0}$ and $M$. There are two processes that
make the system evolute to the other Fock states. Process one is
scattering between a pair of atoms, which can makes the spin
component 1 and -1 jump to 0 and 0, and vice versa, due to spin
dependent interaction. The system evolute to Fock state with the
same $M$. Process two is absorbtion of a $\mu=1(-1)$ photon and then
emission of a $\mu=-1(1)$ photon by an atom. The system evolute to
Fock state with $M\pm2$, i.e. the $z$ direction angular momentum of
the spinor BEC increase(decrease) $2\hbar$.

Assume that the initial photon state is $|1,0\rangle$, i.e. the
initial state of the system is $|N_{0},M;1,0\rangle$. The time
evolution of wave function is given by
\begin{eqnarray}
&&|\Psi(t)\rangle=e^{-iH_{eff}t/\hbar}|N_{0},M;1,0\rangle\notag
\\
&&=\sum_{S',\mu}\sum_{\nu}\sum_{S}|S',\mathcal{M}-\mu;\{\mu\}\rangle\langle
S',\mathcal{M}-\mu;\{\mu\}| e^{-iH_{eff}t/\hbar}\notag
\\
&&|\mathcal{M},\nu\rangle\langle
\mathcal{M},\nu|S,M;\{\mu_{0}\}\rangle \langle
S,M;\{\mu_{0}\}|N_{0},M;\{\mu_{0}\}\rangle \notag_{\{\mu_{0}=1\}}
\\
&&=\sum_{S',\mu}|S',\mathcal{M}-\mu;\{\mu\}\rangle
H^{S',\mu}_{N_{0},\mathcal{M},\{\mu_{0}=1\}}(t)\label{e14}
\end{eqnarray}
where $\mathcal{M}=M+1$. The matrix element between Fock state and
total spin eigenstate $\langle
S,\mathcal{M}-1|N_{0},\mathcal{M}-1\rangle$ is given in reference
\cite{ma08}. After extracting the spin state of the Nth atom from
the total spin eigenstate $|S',\mathcal{M}-\mu;\{\mu\}\rangle$ with
equation (\ref{e6}), we can calculate the evolution of population of
each spin component as
\begin{equation}
P_{\mu_{0},\sigma}=P_{\mu_{0},\sigma}^{\mu=1}+P_{\mu_{0},\sigma}^{\mu=-1}
\label{e15}
\end{equation}
\begin{eqnarray}
P_{\mu_{0},\sigma}^{\mu}&&=\sum_{S}[(A^{N}_{S,\mathcal{M}-\mu,\sigma})^{2}+(B^{N}_{S,\mathcal{M}-\mu,\sigma})^{2}]|H^{S,\mu}_{N_{0},\mathcal{M},\{\mu_{0}=1\}}(t)|^{2}\notag
\\
&&+\sum_{S'}2A^{N}_{S'-1,\mathcal{M}-\mu,\sigma}B^{N}_{S'+1,\mathcal{M}-\mu,\sigma}\notag
\\
&&Re\{H^{S'-1,\mu}_{N_{0},\mathcal{M},\{\mu_{0}=1\}}(t)[H^{S'+1,\mu
}_{N_{0},\mathcal{M},\{\mu_{0}=1\}}(t)]^{*}\} \label{e16}
\end{eqnarray}
where $N-S$ is even and $N-S'$ is odd. The probability of finding
the photon in cavity mode $\{\mu\}$ is given as
$P_{\mu_{0}}^{\mu}=\sum_{\sigma=\pm1,0}P_{\mu_{0},\sigma}^{\mu}$. If
the initial photon state is $|0,1\rangle$, the same equation with
$\mathcal{M}=M-1$ and $\{\mu_{0}=-1\}$ describes the time evolution.
If the initial photon state stay in two cavity modes with the same
probability, the evolution is given by
$P_{\sigma}=(P_{\mu_{0}=1,\sigma}+P_{\mu_{0}=-1,\sigma})/2$. We can
see that the time evolution is decided by symmetric of the atomic
system as well as coupling strength and spin dependent interaction
strength.

\begin{figure}[tbp]
\scalebox{0.6}{\includegraphics{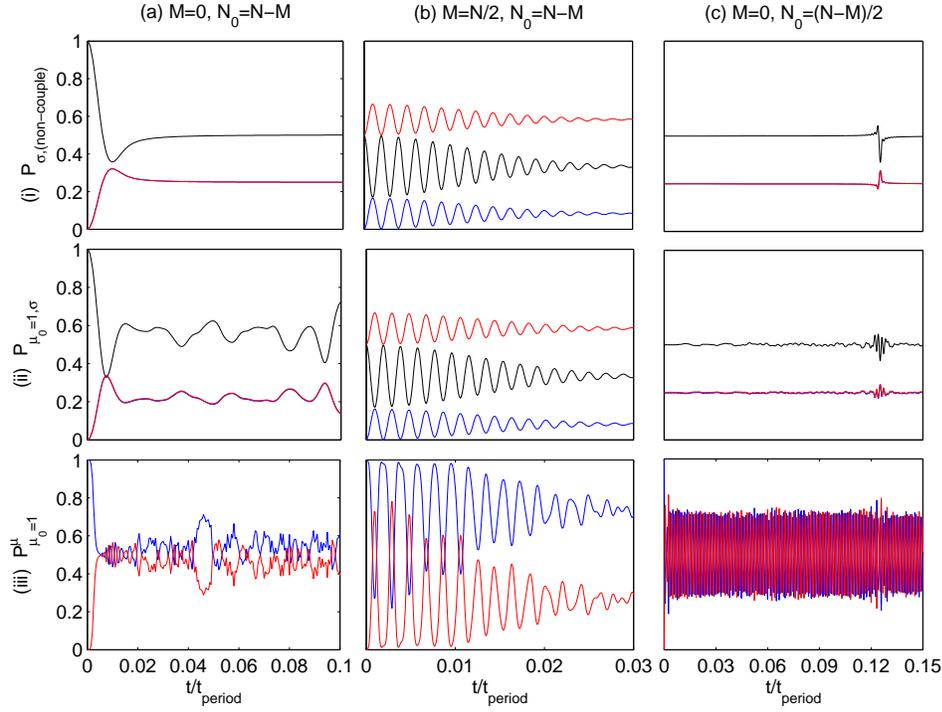}}
\caption{ Column (a) (b) and (c) is numerical result with different
initial condition marked at the subtitle. Atomic number is $N=300$.
Row (i) and (ii) is time evolution of the average population at spin
component $\sigma=0$(black), 1(red), and -1(blue) without and with
coupling between atoms and cavity, respectively. Row (iii) is time
evolution of probability of the photon state at $\mu=1$(blue) and
-1(red) polarization mode.  } \label{fig1}
\end{figure}

Numerical result of 300 $Rb^{87}$ atoms with various initial
condition is shown in figure \ref{fig1}. $J$ is estimated by
Thomas-Fermi approximation as $J=\hbar(\omega_{trap}^{2}/N)^{3/5}/X$
with $X=1.52\times10^{4}$ and $\omega_{trap}=514.5Hz$ being
frequency of the optical trap \cite{11chang05},
$\omega_{e}=2\pi\times 377.107THz=2\omega_{c}$, and
$\Omega_{0}=2\pi\times1MHz$ which is available in experiment
\cite{02manuz05}. The time is normalized by $t_{period}=\pi\hbar/J$,
which is period of time evolution when there is no coupling between
atoms and cavity. From figure \ref{fig1} (a), comparison of (i) and
(ii) shows that time evolution of coupling system has oscillation at
time zone that non-coupling system is stationary. In figure (iii),
the time evolution of probability of photon state at each
polarization mode becomes random and fast oscillating around 0.5
after a short time of smooth relation. Changing to another initial
condition in figure (b), time evolution of coupling and non-coupling
system are almost the same. The photon state keeps oscillation
between two polarization modes and relax to be stable. In figure
(c), it was predicted in a former work that this kind of initial
condition result in a long stable evolution and then a short plus at
time $t_{period}/8$. Figure (ii) shows that the coupling makes the
stable evolution contains a small amplitude oscillation. The
frequency of the oscillation small at early time and keep
increasing, and then become a multi-plus at time $t_{period}/8$. In
figure (iii), it is shown that the photon state oscillates at a very
large frequency between two polarization modes with a quasi-constant
amplitude.

\section{Conclusion and Discussion}

We have model the coupling system of $N$ $Rb^{87}$ atoms and high Q
optical cavity mode by two level approximation, i.e. only the
transition $5^{2}S_{1/2},F=1\longrightarrow5^{2}P_{1/2},F=1$ of D
line is consider. A more accurate model could be including other
transition in D line of $Rb^{87}$. Because the frequency of optical
cavity mode is far off resonance from transition of D line, the
operator of excited state in the Hamiltonian can be removed, which
result in an effective Hamiltonian. The effective Hamiltonian
describe the coupling by two photons process, absorbtion of one
photon from one cavity mode and then emission of one photon to
another cavity mode. The eigenstate, or the dressed spinor BEC
state, is obtained by diagonalizing the Hamiltonian under the basic
set of direct product states, which are consist of total spin
eigenstate and photonic Fock state of non-coupling Hamiltonian. The
FPCs exhibit strong power in deducing the matrix elements of the
Hamiltonian. With the dressed states in hand, the time evolution of
wave function and average population in each spin component is
derived. From the evolution equation (\ref{e14}) (\ref{e15}) and
(\ref{e16}), when the system evolute to the direct product state
with $\{\mu\neq\mu_{0}\}$, the photon evolute from the initial
cavity mode to another cavity mode, and at the same time the spinor
BEC evolute to the state with different $z$ direction angular
momentum. This process result in entanglement of the cavity modes
and the spinor BEC. The numerical result show that the time
evolution with some special initial condition is greatly modified by
the cavity QED coupling. This effect is supposed to be observable
experimentally. The time evolution with some other special initial
condition, the oscillation of probability of finding a photon in
each cavity mode is regular, which could be useful in quantum
entanglement application or exploration.

\bigskip

\begin{acknowledgments}
We appreciate the support from the NSFC under the grants 10574163
and 10674182.
\end{acknowledgments}

\clearpage

\end{document}